\documentclass[11pt,onecolumn]{IEEEtran}
\usepackage[dvips]{graphicx}
\usepackage{epsfig}
\usepackage{amsopn}
\usepackage{verbatim}
\usepackage{amsmath}

\input{prepictex}
\input{postpictex}
\usepackage{color}
\definecolor{turq}{cmyk}{0.25,0,0.5,0}
\definecolor{mycy}{cmyk}{0.2,0,0,0}
\definecolor{mymag}{cmyk}{0,0.3,0,0}
\definecolor{brown}{rgb}{0.8,0.6,0.1}

\newcommand{\cw}{\mbox{\boldmath $c$}}

\newcommand{\be}{\begin{equation}}
\newcommand{\ee}{\end{equation}}
\newcommand{\ew}{\mbox{\boldmath $e$}}

\newcommand{\rf}[1]{\mbox{(\ref{#1})}}

\newcommand{\bi}{\begin{itemize}}
\newcommand{\ei}{\end{itemize}}
\newcommand{\TR}{\mbox{tr}}
\newcommand{\Prob}{\mbox{Pr}}
\newcommand{\bd}[1]{\mbox{\boldmath $#1$}}

\newcommand{\ho}[1]{\mbox{\hollow #1}}
\newcommand{\df}{\stackrel{\mbox{\scriptsize def}}{=}}
\newcommand{\dg}{\dagger}

\newcommand{\sh}{\mbox{$S_{\ell}$}}

\begin{document}

\title{On Interleaving Techniques for MIMO Channels and Limitations of Bit Interleaved Coded Modulation}

\author{D.\ Mihai Ionescu,~\IEEEmembership{Senior Member,~IEEE},
D\~ung~N.~Doan,~and  Steven D.\ Gray,~\IEEEmembership{Senior
Member,~IEEE}
\thanks{Dumitru Mihai Ionescu is with Nokia Research Center, 12278 Scripps Summit Dr., San Diego, CA 92131, USA (e-mails:
michael.ionescu@nokia.com, dmicmic@ieee.org).}
\thanks{D\~ung~N.~Doan was with the Department of Electrical and
Computer Engineering, Texas A\&M University, and with Nokia
Research Center. He is now with Qualcomm Inc., San Diego, CA
(e-mail:~ddoan@qualcomm.com).} \thanks{Steven D.\ Gray was with
Nokia Research Center, 6000 Connection Dr., Irving TX 75039; he is
now with Intel.}}

\date{}
\font \hollow=msbm10 at 12pt \maketitle

\begin{abstract} It is shown that while the  mutual information
curves for coded modulation (CM) and bit interleaved coded
modulation (BICM) overlap in the case of a single input single
output channel, the same is not true in multiple input multiple
output (MIMO) channels. A method for mitigating fading in the
presence of multiple transmit antennas, named coordinate
interleaving (CI), is presented as a generalization of component
interleaving for a single transmit antenna. The extent of any
advantages of CI over BICM, relative to CM, is analyzed from a
mutual information perspective; the analysis is based on an
equivalent parallel channel model for CI. Several expressions for
mutual information in the presence of CI and multiple transmit and
receive antennas are derived. Results show that CI gives higher
mutual information compared to that of BICM if proper signal
mappings are used.  Effects like constellation rotation in the
presence of CI are also considered and illustrated; it is shown
that constellation rotation can increase the constrained capacity.
\end{abstract}


\section{Introduction}
Mitigating the effects of fading on the decoding of forward error
correction codes has been studied for single transmit antenna
configurations, and revolves around some form of interleaving.
Traditionally, interleaving was performed directly on the coded
bits present at the encoder's output. Given that coded modulation
(CM) \cite{ung:cha} is optimal, the recognition of the fact that
modulation and  coding are no longer necessarily combined when
interleaving an encoder's output---i.e., prior to mapping the
(interleaved) bits to complex (channel) alphabet symbols---has
naturally led researchers \cite{cai:bit} to study capacity
limitations of bit interleaved coded modulation (BICM) schemes;
see \cite{cai:bit} and references therein for definition of BICM.
In the case of single transmit antenna configurations, the
conclusion was that mutual information remains essentially the
same---provided that Gray mapping is used to map the interleaved
coded bits to modulator's points. This reinforced a potential
advantage of BICM, i.e., flexibility from the perspective of
adaptive coding and modulation schemes. Other aspects of the study
in \cite{cai:bit} related to the cut-off rate---but this concept
has been deemed irrelevant with the advance of iterative decoding
and concatenated coding schemes (e.g., turbo codes). As a result,
BICM was deemed desirable, and was employed in standards for  CDMA
\cite{3GPP2:add} and satellite systems that involve one transmit
antenna.

In parallel, an alternative to separating coding from modulation
was studied for single transmit antenna configurations, and named
coordinate (or component) interleaving \cite{jel:des}.  In that
approach, the real and imaginary coordinates were interleaved
separately, and diversity was derived from the encoder's trellis
(minimum Hamming distance); there was no difficulty reading the
interleaved coordinates, because complex symbols were transmitted
from only one transmit antenna, and the orthogonal in-phase and
quadrature components could  be naturally and separately
detected\footnote{Assuming that imperfections such as leakage
between in-phase and quadrature (due to synchronization errors)
are absent.}. The idea of interleaving the real coordinates, or
components, of points from multidimensional constellations
 embedded in some Cartesian product of the complex field \ho{C} was discussed
 in \cite{bou:sig}; therein, an algebraic number theoretic analysis was pursued
 to support the conclusion that coordinate interleaving (CI) together with
 constellation rotation can increase diversity---separately from any redundancy
 scheme such as forward error correction coding. The diversity was quantified
 in terms of a coordinate-wise Hamming distance. In \cite{fur:rot} the authors
 restricted the MIMO channel to a two transmit antenna configuration, and
 relied on constellation rotation along with an orthogonal space-time
 diversity scheme \cite{ala:sim} to increase diversity by increasing
 the coordinate-wise Hamming distance;  coordinates of a rotated complex
 constellation were first interleaved then mapped to  orthogonal
 space-time block code  matrices for two transmit antennas  \cite{ala:sim}.
 By allowing the symbols from the two transmit antennas to be
 easily separable at the receive antenna due to orthogonality,
 orthogonal space time block codes allow the approach in
 \cite{jel:des} to be extended to two transmit antennas.

The problem is more complicated when multiple transmit antennas are employed,
as coordinates can no longer be observed independently from each other due to
the superposition of all transmitted complex symbols at any  receive antenna.
A version of BICM for multiple input multiple output (MIMO) channels
was examined in \cite{bou:bit,ber:spa}.

A more general scheme for CI in the presence of multiple transmit
antennas was reported in \cite{ion:ful}, and summarized in
Section~\ref{ci} below; while the exemplary illustration in
\cite{ion:ful} was based on a two transmit antenna configuration,
it can be applied to any number of transmit antennas. In addition,
it can be employed simultaneously with a coding redundancy scheme
(concatenated or not), does not require separating coding from
modulation, and does not rely on constellation rotation---although
rotation is not precluded; it is easily recognized to differ from
channel interleaving, which would interleave blocks of (all)
complex symbols to be transmitted from all transmit antennas
during one MIMO channel use. This CI scheme offers a means to
mitigate diversity in MIMO channels, which generalizes the single
antenna concept in a non-obvious manner. The minimum
coordinate-wise Hamming distance can be naturally controlled in
the design of the forward error correcting scheme; in essence,
after coding and modulation (possibly combined) the real and
imaginary coordinates from the complex symbols of all transmit
antennas are interleaved over an arbitrary block length, e.g.\ a
frame, and then transmitted as new coordinates in the respective
MIMO configuration. Further considerations may be possible using
the algebraic number theoretic formalism undertaken in
\cite{bou:sig}. Depending on the original complex constellation(s)
used on the transmit antennas, the complex channel alphabet after
CI may change over the wireless channel. Moreover, the coordinate
interleaver  can be viewed as an interleaver in a serially
concatenated scheme, thus enabling iterations between decoding and
detection in a manner similar to turbo decoding; this is also
described in a separate work \cite{ion:ful}. This paper is not
concerned with the details of combining coding redundancy and CI,
the latter briefly reviewed in Section~\ref{ci}; rather, the
impact of CI on diversity, via the minimum coordinate-wise Hamming
distance, motivates one to investigate mutual information limits.
This is the main purpose of this manuscript, which sets out to
determine the extent of any advantages of CI over BICM, relative
to CM, from a mutual information perspective. It should be noted
that in this paper we are dealing with capacity limits for finite
alphabet constellations (constrained constellations)---rather than
the generally used {\it ergodic} capacity, which is often
associated with the case of continuous alphabets.

Section~\ref{ci} reviews the mechanism by which CI affects
diversity in a MIMO system. Section~\ref{mutinfCMandBICMwMIMO}
examines the mutual information with finite size alphabets in the
single transmit antenna case \cite{cai:bit}, as well as in several
MIMO configurations; the numerical results  illustrate that the
fact that  mutual information curves for CM and BICM overlap in
the case of a single transmit antenna ought not to be taken for
granted, as a significant gap occurs in MIMO channels.
Section~\ref{mici} introduces an equivalent parallel channel model
for CI, and derives several expressions for mutual information in
the presence of CI and multiple transmit antennas.
Section~\ref{constellnoninvCI} examines effects like constellation
rotation in the presence of CI. Conclusions are drawn in
Section~\ref{concl}.

\section{Coordinate Interleaving for MIMO Channels -- Diversity Perspective}
\label{ci}
Consider a MIMO  scenario where the number of transmit and, respectively,
receive antennas are $N$ and $M$. Let the complex symbols transmitted from the
$N$ transmit antennas during $l$ uses of the MIMO channel be represented as a
long vector, whose entries are superscripted by the transmit antenna index,
and subscripted by the channel use index. Conditioned on knowledge of the
channel state information (CSI), the probability of transmitting
\begin{eqnarray}
\lefteqn{\ew} & \lefteqn{=} & \left[ \begin{array}{@{\,}c@{\,}c@{\,}c@{\,}c@{\,}c@{\,}c@{\,}c@{\,}c@{\,}c@{\,}c@{\,}c@{\,}}
  e_0^{(1)}& e_0^{(2)}& \ldots & e_0^{(N)}& e_{1}^{(1)} &  \ldots & e_{1}^{(N)} & \ldots & e_{l-1}^{(1)} & \ldots & e_{l-1}^{(N)}
\end{array}
\right]^{T} \nonumber
\end{eqnarray}
and deciding in favor of
\begin{eqnarray}
\lefteqn{\cw} & \lefteqn{=} & \left[ \begin{array}{@{\,}c@{\,}c@{\,}c@{\,}c@{\,}c@{\,}c@{\,}c@{\,}c@{\,}c@{\,}c@{\,}c@{\,}}
  c_0^{(1)}& c_0^{(2)}& \ldots & c_0^{(N)}& c_{1}^{(1)} &  \ldots & c_{1}^{(N)} & \ldots & c_{l-1}^{(1)} & \ldots & c_{l-1}^{(N)}
\end{array}
\right]^{T} \nonumber
\end{eqnarray}
at the maximum likelihood decoder is bounded as below
\begin{equation}
\Prob \{\cw \!\mapsto \!\ew | \alpha_{ij}[k], i=1, \ldots, N, j=1, \ldots, M, k=0, \ldots, l-1\} \nonumber \\ \leq \exp(-d^2_E(\ew, \cw) E_s/4 N_0),
\end{equation}
where $\alpha_{i,j}(\cdot)$ are the
 channel  coefficients between
transmit antenna $i$  and receive antenna $j$, with
$E\{|\alpha_{i,j}|^2\}=1$, $\forall i,j$. The key parameter is
clearly \be d^2_E(\ew,\cw)=\sum_{j=1}^M \sum_{k=0}^{l-1}\left|
\sum_{i=1}^N \alpha_{i,j}[k](c_k^{(i)}-e_k^{(i)})\right|^2. \ee As
is well-known \cite{tar:spa} \be d^2_E(\ew,\cw)=\sum_{j=1}^M
\sum_{k=0}^{l-1}\bd{\Omega}_j[k] \bd{C}[k]\bd{\Omega}_j^\dg [k],
\ee where \be \bd{\Omega}_j[k]=\left[ \alpha_{1,j}[k], \ldots,
\alpha_{N,j}[k] \right], \ee \be
\bd{C}[k]=(\bd{c}_k-\bd{e}_k)(\bd{c}_k-\bd{e}_k)^\dg=\left[
\begin{array}{@{\,}c}
c_k^{(1)}-e_k^{(1)}\\c_k^{(2)}-e_k^{(2)}\\ \vdots \\c_k^{(N)}-e_k^{(N)}
\end{array}
\right] \left[ \left(c_k^{(1)}-e_k^{(1)}\right)^*, \ldots,
\left(c_k^{(N)}-e_k^{(N)}\right)^*\right], \ee where the
superscript $\ast$ indicates complex conjugation. Since
$\bd{C}[k]$ is Hermitian, it admits the singular value
decomposition (SVD) \be \bd{C}[k]=\bd{V}[k] \bd{D}[k] \bd{V}^{\rm
H}[k] \ee where the superscript `H' indicates a Hermitian
operation (complex conjugated transposition). Denote by
$D_{ii}[k]$, $1\leq i \leq N$, the diagonal elements of
$\bd{D}[k]$, which is diagonal per SVD transform. The vector
$\bd{\Omega}_j[k]$ of relevant channel coefficients (to receive
antenna $j$) is transformed by virtue of the SVD into \be \left[
\beta_{1,j}[k], \ldots, \beta_{N,j}[k] \right]=\bd{\Omega}_j[k]
\bd{V}[k]. \ee Because $\bd{V}[k]$ is unitary, the independent
complex Gaussian random variables (r.v.s) $\alpha_{1,j}[k],
\ldots, \alpha_{N,j}[k]$ are transformed into a new set of $N$
i.i.d.\ r.v.s. In other words, there exists an equivalent set of
channels $\beta_{1,j}[k], \ldots, \beta_{N,j}[k]$ that
characterizes the transmission. Therefore, for each channel use
$k$, and each receive antenna $j$, \be \bd{\Omega}_j[k]
\bd{C}[k]\bd{\Omega}_j^\dg [k]=\sum_{i=1}^N \left|
\beta_{i,j}[k]\right|^2 D_{ii}[k]. \ee By definition, $\bd{C}[k]$
has rank 1 (that is, if the set
$\begin{array}{@{\,}c@{\,}c@{\,}c@{\,}c@{\,}} c_k^{(1)}&
c_k^{(2)}& \ldots & c_k^{(N)}
\end{array}$ is different from $\begin{array}{@{\,}c@{\,}c@{\,}c@{\,}c@{\,}} e_k^{(1)}& e_k^{(2)}& \ldots & e_k^{(N)} \end{array}$); thereby, exactly one value among $D_{11}[k] \ldots D_{NN}[k]$, be it $D_{i_0 i_0}[k]$, is nonzero. The nonzero value must necessarily equal the trace of $\bd{C}[k]$, which in turn equals obviously
\be \TR \bd{C}[k]=\sum_{i=1}^N |c_k^{(i)}-e_k^{(i)}|^2 =\|
\bd{c}_k - \bd{e}_k\|^2. \ee Consequently, the key parameter
$d_E^2(\bd{e},\bd{c})$ reduces to \be \bd{\Omega}_j[k]
\bd{C}[k]\bd{\Omega}_j^\dg [k]=\left| \beta_{i_0, j}[k]\right|^2
D_{i_0 i_0}[k] = \left| \beta_{i_0, j}[k]\right|^2 \cdot
\sum_{i=1}^N |c_k^{(i)}-e_k^{(i)}|^2. \label{1fade} \ee Clearly,
{\em there exists an equivalent set of independent complex
Gaussian channels derived from the original set of independent
complex Gaussian channels, and exactly one of them affects all of
the coordinates of the multidimensional point
$\begin{array}{@{\,}c@{\,}c@{\,}c@{\,}c@{\,}} (c_k^{(1)}&
c_k^{(2)}& \ldots & c_k^{(N)}) \end{array}$ transmitted during
channel use $k$}; this means that if the nonzero equivalent
channel $\beta_{i_0, j}[k]$ fades, it will affect all $2N$
coordinates transmitted during the $k$th channel use. The essence
of CI is summarized by
\newtheorem{coordinterleave}{Theorem}\begin{coordinterleave}
There exists an equivalent set of independent complex Gaussian
channels derived from $\alpha_{1j}[k]\!,\! \ldots \!,
\!\alpha_{Nj}[k]$, such that exactly one of them affects all
(real/imaginary) coordinates of a transmitted multidimensional
point $(c_k^{(1)}, \ldots, c_k^{(N)})\in \!\!\ho{R}^{2 N}$, via
$\bd{\Omega}_j[t] \bd{C}[t]\bd{\Omega}_j^\dg [t]\!=\!|\beta_{i_0,
j}(t)|^2 \sum_{i=1}^{N}\!\!\left[\left(c_{t,{\rm
I}}^{(i)}-e_{t,{\rm I}}^{(i)}\right)^2\!\!+\!\left(c_{t,{\rm
Q}}^{(i)}-e_{t,{\rm Q}}^{(i)}\right)^2\right]$.
\label{coordinterleave}
\end{coordinterleave}

Theorem~\ref{coordinterleave} implies that, when using multiple
transmit antennas and CM possibly over non-binary fields, with or
without puncturing, it is desirable to: \begin{enumerate} \item
Interleave the coordinates of transmitted multidimensional
constellation points in order to enable, and render meaningful, a
Hamming distance with respect to coordinates---rather than complex
symbols; and, \item   Design codes for multiple transmit antennas
that can maximize the minimum coordinate-wise (as opposed to
complex-symbol-wise per current state of the art \cite{tar:spa})
Hamming distance between codewords, in order to derive the most
diversity from CI\end{enumerate}

Note that, in general, CI is different from
BICM \cite{cai:bit} (see also
Example~\ref{ex1}), and does {\em not} preclude (or destroy) the
concept of CM (via signal-space coding); this is so because
CI operates on the (real-valued) coordinates
of the complex values from the complex modulator alphabet, rather
than operating on the coded bits, prior to the modulator
\cite[Fig.~1]{cai:bit}.

It is possible to design space-time codes for CI that exhibit
diversity exceeding that of existing codes; this, however, is the
scope of a separate work \cite{ion:ful}. The sequel is concerned
with characterizing CI from a mutual information perspective.

\section{Mutual information limits for CM and  BICM: Is BICM
Efficient in MIMO Channels?} \label{mutinfCMandBICMwMIMO} 
\begin{figure}[htbp] \begin{center} \epsfig{file=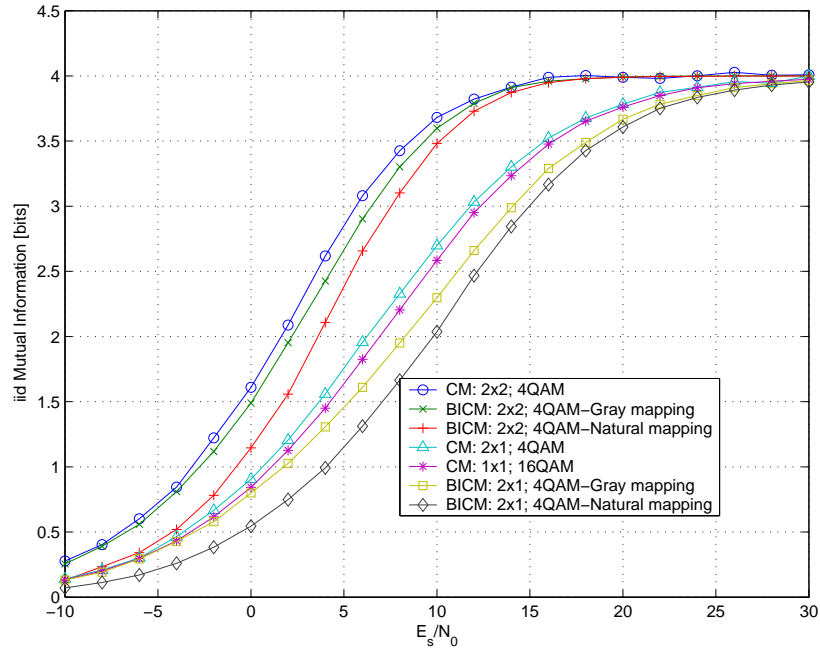, width=0.6\textwidth} \caption{Mutual information for CM vs.\ BICM \mbox{\protect\cite{cai:bit}}, with  finite 4PSK modulation alphabets, two transmit antennas and either 1 or 2 receive antennas, and memoryless  fading channels. The mutual information for CM was computed based on the model and approach from \mbox{\protect\cite{cai:bit}}, using the computational tool described in \mbox{\protect\cite{arn:inf}}, based on the BCJR algorithm.} \label{fig:mutinf4PSKdoan} \end{center} \end{figure}

If the
model in \cite{cai:bit} is used to characterize mutual information
limits for CM and BICM schemes that involve $N>1$  transmit
antennas (only scenario $N=1$ was discussed in \cite{cai:bit}) the
curves in Fig.~\ref{fig:mutinf4PSKdoan} are obtained; the
numerical results were computed by an extension to MIMO channels
of the analysis (and numerical approach) from \cite{cai:bit}. The
curves in Fig.~\ref{fig:mutinf4PSKdoan} indicate a fundamental
difference between the information rate limits in scenarios that
involve finite channel alphabets, and more than one
transmit/receive antennas---even for the Gray mapping case, which
in \cite{cai:bit} is conjectured to maximize the mutual
information of bit-interleaved schemes. (It is possible to view CM
as a concatenation between a code and a mapper, {\em provided that
the mapper is derived from some set partitioning labelling}.)
While it can be rigorously proven that CI does not change the {\em
ergodic} channel capacity, the model in \cite{cai:bit} predicts a
gap between mutual information curves for BICM relative to pure CM
(due, in part, to the finite size channel alphabet; see, however,
the footnote in \cite[p.\ 931]{cai:bit}). The relevant mutual
information curves are illustrated in
Fig.~\ref{fig:mutinf4PSKdoan}; with $N=2$  transmit antennas and
$M=1$ receive antenna, the gap is significant, but narrows when
$M=2$.

Note that in CI schemes the role of iterating between decoding and
detection is to circumvent maximum likelihood detection; the
presence of an interleaver does not always preclude maximum
likelihood---e.g., a symbol interleaver in a CM scheme does not.
It should be noted that if iterative demodulation and decoding is
used for CI (same as for BICM), it is possible to approach the
i.i.d. CM capacity limit, since iterations improve the quality of
the soft information being exchanged between the demodulator and
decoder. However, this paper attempts to quantify  only the case
of non-iterative demodulation/decoding.

\begin{figure}[htbp] \begin{center} \epsfig{file=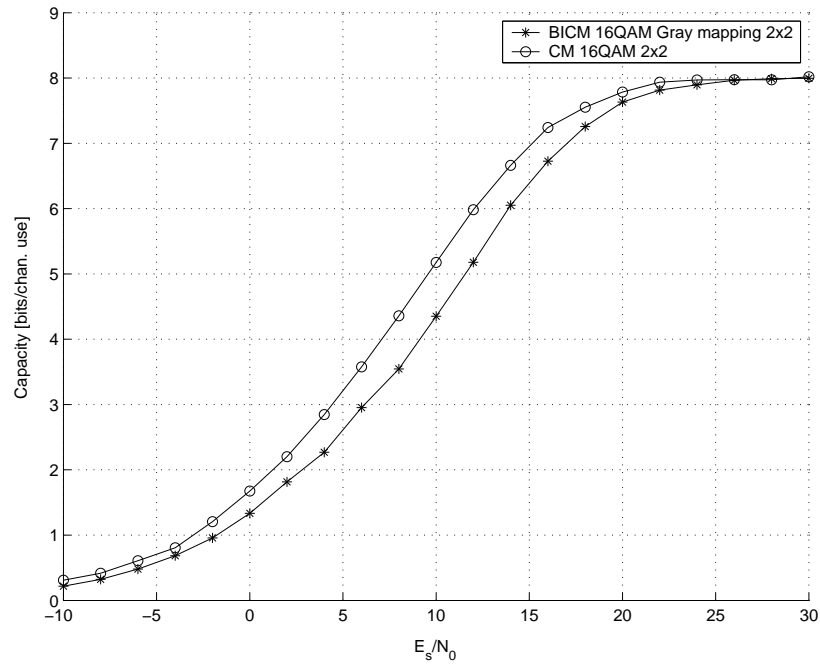, width=0.6\textwidth} \caption{Mutual information for CM vs.\ BICM \mbox{\protect\cite{cai:bit}}, with  finite 16QAM modulation alphabets, two transmit antennas and 2 receive antennas, in memoryless fading channels; note increased gap. The mutual information for CM was computed based on the model and approach from \mbox{\protect\cite{cai:bit}}, using the computational tool described in \mbox{\protect\cite{arn:inf}}, based on the BCJR algorithm.} \label{fig:mutinf16QAMdoan} \end{center} \end{figure}

To further illustrate the undesirable effect of BICM, Fig.~\ref{fig:mutinf16QAMdoan} shows the relevant mutual information curves when $N=M=2$, and 16QAM is used on the individual transmit antennas.

From the above 2 figures, we see that BICM suffers a big loss in
mutual information compared to CM in MIMO channels. This, together
with the diversity advantage of CI shown in the previous section
motivate us to investigate the CI in a mutual information
perspective. Results will show that CI outperforms BICM in terms
of mutual information and narrows the gap relative to CM. The
following section is a derivation of mutual information for a CI
scheme in MIMO channels.

\section{Mutual Information for Coordinate Interleaved Schemes in
MIMO Rayleigh Fading Channels} \label{mici}

As a result of coding and modulation, $2 N$ (in-phase and
quadrature)  coordinates are generated during any channel use for
transmission from the $N$ transmit antennas; coordinates  are
collected from all MIMO channel uses (over a long enough frame),
and  interleaved in such a manner as to insure that any $2 N$
coordinates---which, in the absence of CI, would be transmitted
from the $N$ transmit antennas during the same MIMO channel
use---are transmitted (as in-phase and quadrature components) from
different antennas, during different MIMO channel uses; i.e., each
of the $2 N$ coordinates will experience independent  fading
relative to each other [but the same fading as some other $2 N -1$
coordinates along with which it is eventually transmitted during
some channel use, per eq.\  \rf{1fade}]. Specifically, let the
number of MIMO channel uses in one coordinate interleaved frame be
$l$; the coordinate interleaver size is then $2 N l$. Denote by
$\bd{x}_{k'}$ the $2N$ dimensional symbol formed by the $2N$
coordinates that are to be transmitted, {\em post}  CI, from the
$N$ transmit antennas during the $k'$-th  MIMO channel use, $k'=
0, \ldots, l-1$: \be \textstyle { \bd{x}_{k'} \in {\cal X}\df
\{\bd{x} | \bd{x}=[x_0, \ldots, x_{2N-1}]^{\rm T} \sim [x(1)\ldots
x(N)]^{\rm T}\}, }\label{nalabel} \ee where $\sim$ denotes a known
isomorphism between real and complex vectors, and \be
x(m+1)=x_{2m}j +x_{2m+1}\in {\cal M}, \ \   m=0, \ldots, N-1, \ee
is a complex symbol, from some constellation $\cal M$ (see below),
to be sent from the $(m+1)$-st transmit antenna; the constellation
$\cal M$ will arise as the Cartesian product of the in-phase and
quadrature alphabets {\em post}  CI. (The resulting constellation
mapping after CI, $\cal M$, will be the same with  the original
mapping for square QAM modulation; and different for general PSK
modulation.) Further, denote by $\{c_k\}_{0}^{2 N l-1}$,
respectively $\{\kappa_k\}_{0}^{2 N l-1}$, the coordinates before
and after interleaving. A coordinate $2N$-tuple $\{\kappa_k\}_{2 N
p}^{2 N p+2N-1}$, $p \in {\ho N}$, will be said to form an
$N$-antenna label, because they would all form the $2N$
coordinates to be sent from the $N$ antennas during the same
channel use; more specifically, the coordinates whose index is
congruent modulo 2 with 0, respectively 1, will form the
quadrature and in-phase components.

In order to model the CI operation,  let $c_k$ be permuted to
$\kappa_{\pi_{0} (k) 2 N+ \pi_{1}(k)}$, i.e., the coordinate
interleaver permutation will be $\pi : k \mapsto \left(\pi_{0}
(k),  \pi_{1}(k)\right)$, $\pi_{1}(k)=0,1,  \ldots, 2N -1$,
$\pi_{0}(k)=0,1,  \ldots, l -1$. That is, $c_k$ is transmitted
during the $\pi_{0}(k)$-th MIMO channel use, as the
$\pi_{1}(k)$-th component of the $N$-antenna label of
$\bd{x}_{k'}|_{k'=\pi_{0} (k)}$. The receiver has to detect the
$i$-th component of $\bd{x}_{k'}$ and unpermute it to obtain
$c_{\pi^{-1}(k', i)}$. Note that the component pairs $x_{2m},
x_{2m+1}$ are affected by the same fade, namely the channel fading
coefficient on the link between transmit antenna $m+1$ and the
relevant  receive antenna.

\begin{figure}[bt]
\scalebox{0.5}{\input{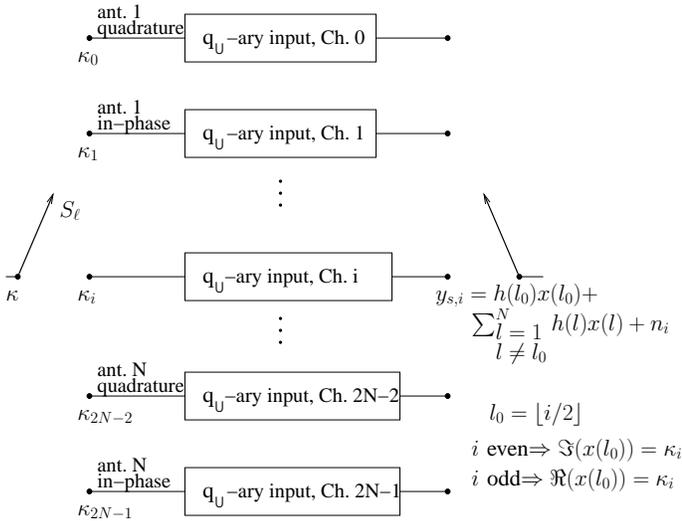}} \caption{Equivalent
parallel channel model for computing  mutual information in the
presence of  CI---assuming ideal (coordinate) interleaving, $N$
transmit antennas; depicted channel  conveys one $q_\cup$-ary
r.v., carried in a random position of an $N$-antenna label. The
total mutual information with CI will be $2 N$ times the capacity
of this channel; $y_{s,i}$ is the observation of the $i$-th
channel at the $s$-th  receive antenna. $h(l)$ represents the
channel coefficient from the $l$-th transmit antenna to the $s$-th
receive antenna; $x(l)$ is the complex value, from some complex
alphabet $\cal M$, transmitted  from the $l$-th transmit antenna.
The complex constellation $\cal M$ arises as the Cartesian product
of the quadrature and in-phase alphabets {\em post}  component
interleaving; $q_\cup$ is the cardinality of both the in-phase and
the quadrature alphabets {\em post} component interleaving---each
coordinate alphabet being, in turn, the union of the in-phase and
quadrature alphabets {\em prior to} component
interleaving.}\label{parchann}
\end{figure}

As long as  the coordinate interleaver has sufficient depth (to
insure exposure to independent fades), the index $\pi_{0}(k)$ of
the MIMO channel use that conveys, after interleaving, a
(pre-interleaving) coordinate $c_k$   becomes irrelevant from the
perspective of computing the  mutual information; further,
coordinate $c_k$ appears as a random element $\pi_{1}(k)$ of some
$N$-antenna label  (a $2N$-dimensional point), whose components
are detected individually---after having been affected by a common
fading coefficient (see eq.\  \rf{1fade}). The random position in
an $N$-antenna label can be modeled by a switch \sh, and the
equivalent parallel model is shown in Fig.~\ref{parchann}.

The complex constellation $\cal M$ mentioned above arises as the
Cartesian product of the quadrature and in-phase alphabets {\em
post}  component interleaving; in order to further define the
formalism, assume for simplicity that prior to CI the modulator
employs the same complex constellation $\cal Q$ on each of the $N$
transmit antennas. Let ${\cal Q}_I$ and ${\cal Q}_Q$ be,
respectively, the alphabets pertaining to the in-phase and
quadrature coordinates of points from $\cal Q$ {\em prior} to CI.
Since CI mixes the in-phase and quadrature coordinates, each
coordinate post interleaving will belong to an alphabet $Q_\cup$
that is the union of the alphabets ${\cal Q}_I$ and ${\cal Q}_Q$,
${\cal Q}_\cup \df {\cal Q}_I \cup {\cal Q}_Q$. Therefore, let \be
q_\cup  \df |{\cal Q}_\cup| = |{\cal Q}_I \cup {\cal Q}_Q| \ee be
the cardinality of both the in-phase and the quadrature alphabets
{\em post} component interleaving---each coordinate alphabet
being, in turn, the union of the in-phase and quadrature alphabets
{\em prior to} component interleaving. Then the effective
(post-interleaving) complex constellation associated with any
transmit antenna becomes \be {\cal M}=({\cal Q}_I \cup {\cal
Q}_Q)\times ({\cal Q}_I \cup {\cal Q}_Q),
\label{perantciconstell}\ee and the MIMO constellation post CI,
$\cal X$, is the $N$-fold Cartesian product of $\cal M$ with
itself, \be {\cal X}= {\cal M}^N. \label{ciconstell} \ee
Finally,
the operation of mapping  real (post-interleaving) coordinates
$\kappa_k$ to multidimensional points is defined naturally, in
that  a coordinate $\kappa_k$ simply becomes a component of some
multidimensional point $\bd{x} \in {\cal X}$, and its label;
unlike in BICM \cite{cai:bit}, where Gray mapping has been
conjectured to be optimal, the mapping of coordinates to
$N$-antenna labels is unambiguous (no multiple choices).

The problem  then becomes calculating the mutual information
between pre-interleaving  $2 N$-tuples $\left[ c_0,  c_1, \ldots ,
c_{2N-1} \right]^{\rm T}$ and the receiver observations at the $2
N$ different MIMO channel uses\footnote{Per assumption at
beginning of Section~\ref{mici}.} that convey, post component
interleaving, those coordinates; clearly this model involves
buffering. If CI is such as to insure that each uninterleaved
coordinate is sent during independent MIMO channel uses, than the
desired mutual information is $2 N$ times the mutual information
of an abstract  channel conveying a single, $q_\cup$-ary,
post-interleaving coordinate, i.e.\ between an interleaved
coordinate  and the observations---by the $M$ receive
antennas---of the multidimensional point from ${\cal M}^N$ that
carries the corresponding interleaved coordinate. The assumption
of ideal interleaving reduces, thereby,  the relevant problem  to
the random switch model in Fig.~\ref{parchann}. The auxiliary
switch \sh \ will be eliminated by averaging over its possible
realizations from $\{0, \ldots , 2N-1\}$; this will require one to
compute the mutual information of the $i$-th channel, i.e.\
between $\kappa_i$ and $\bd{y}_i=[y_{1,i} \ldots y_{M,i}]^{\rm T}$
in Fig.\ \ref{parchann}, where $y_{s,i}$ is the complex
observation by receive antenna $s$ of the output of the $i$-th
channel,  $i$ is a realization of the  r.v.\  \sh, and the
$\kappa_i$-s are i.i.d.\footnote{The i.i.d.\ assumption for the
$\kappa_i$-s does not hold, in general,  for constellations that
increase their size after CI.}. Therefore, the effect of CI  is
modeled by the operation  of the switch \sh, and by the assumption
of independence between MIMO channel uses.

\newtheorem{ex1}{Example}
\begin{ex1} \label{ex1}
By visual inspection, one easily and rigorously verifies that when $\cal Q$ is an unrotated 4QAM constellation, $ {\cal Q}_I = {\cal Q}_Q ={\cal Q}_\cup$, ${\cal M}= {\cal Q}$, $q_\cup = 2$---i.e.\ the inputs to the parallel channels in Fig.~\ref{parchann} are binary r.v.s, like in BICM; on another hand the BICM channel input is an $m$-tuple with $m=N \log_2 |{\cal Q}| = 2 N$. Thereby, Fig.~\ref{parchann} reduces to the equivalent BICM model from \cite[Fig.~3]{cai:bit} that corresponds to using a 4QAM constellation on each transmit antenna (although \cite{cai:bit} restricts itself to a single transmit antenna, the BICM model from \cite{cai:bit} is directly generalized to $N>1$ with the additional averaging in the form of marginal probabilities, as discussed in Section~\ref{cohvsnon}). Consequently, BICM and CI are equivalent in the 4PSK case, for any  number of transmit antennas $N$. This proves in part (i.e.\ for the 4PSK scenario) Biglieri {\em et al}.'s  conjecture made in \cite{cai:bit} that Gray mapping is optimal for BICM. \hfill $\diamondsuit$
\end{ex1}

In the context defined above, the sequel addresses the scenario
when the receiver has perfect or partial channel state
information, or no channel state information---and, of course, the
information necessary to  unpermute the coordinates; the latter
translates in information about the switch \sh. In general, as it
will turn out below, computing the desired mutual information in
the presence or absence of channel state information reduces to
computing $I(\kappa;\bd{y},\bd{\theta}, \sh=i)$ or
$I(\kappa;\bd{y}, \sh=i)$, where $\bd{\theta}$ represents a
(generalized) channel state parameter, i.e.\ a vector parameter on
which the transition between channel input and output depends. The
mutual information  $I(\kappa;\bd{y},\bd{\theta}, \sh=i)$ will be
derived  below in order to reveal the computational approach for
numerically computing it; the modifications necessary to obtain
$I(\kappa;\bd{y}, \sh=i)$ will be straightforward.

\subsection{Problem setting and notations}
Lower case symbols will denote scalar quantities (real or complex), boldface symbols will denote vectors, and underlined boldface symbols will denote sequences of scalars or of vectors.

In general, the channel input at channel use $k$, denoted generically  $\bd{x}_k$, can be viewed as a point (vector) from some real $N_1$ dimensional space ${\ho R}^{N_1}$---or, for even $N_1 \geq 2$, from a complex  $N_1/2$ dimensional space ${\ho C}^{N_1/2}$; the complex vector entries can be further limited to some complex (i.e., real two dimensional) alphabet $\cal Q$, in which case the overall  complex  $N_1/2$ dimensional alphabet, denoted $\cal X$, has finite cardinality $|{\cal X}|\df M_0$, often verifying $M_0=2^m$. In the case of CM for MIMO channels the  channel input relevant to computing the mutual information is a complex $N$ dimensional point (vector), whereby $N_1=2 N$---see \cite{cai:bit}, albeit \cite{cai:bit} only addressed the single antenna scenario $N=M=1$); for BICM  the relevant channel input is a bit $m$-tuple \cite{cai:bit}, i.e.\ a point from  ${\ho R}^{m}$, and a parallel channel model  \cite[Fig.~3]{cai:bit} reduces it to a single bit, i.e.\ a point from $\{0,1\}\in {\ho R}$, whereby $N_1 =1$. In the CI case, the relevant channel input is a point in ${\ho R}^{2N}$, and the parallel channel model from Fig.~\ref{parchann} reduces it to a single dimensional real point from ${\cal Q}_\cup \df {\cal Q}_I \cup {\cal Q}_Q$.

The output $\bd{y}$ of the channel, generally from some ${\ho C}^{N_2}$ space, can be viewed for the MIMO problem at hand as a point (vector) from a complex $M$ dimensional space, with $M$ being  the number of receive antennas, whereby $N_2= M$; this is true both for the  aggregate MIMO channel and for any one of the parallel channels in Fig.~\ref{parchann}.

In order to correctly identify  the state vector as a point from
some complex space ${\ho C}^{N'}$, consider the $i$-th  parallel
channel in Fig.~\ref{parchann}, and note that the channel output
is a collection of $M$ receive antenna observables; even when the
channel from any transmit antenna to any receive antenna is
memoryless, the observable at receive antenna $s$ depends not only
on the current channel coefficients and on coordinate $\kappa_i$
(during the current MIMO channel use), but also on the remaining
coordinates---other than $\kappa_i$---arriving at the receive
antenna $s$ on the other channels during the current MIMO channel
use, and acting as interference. Thereby the state vector for the
$i$-th subchannel is

\begin{eqnarray} \lefteqn{\protect\bd{\theta}_i   =  \left[ h_{1,1}, \ldots , h_{1,M}, \ldots , h_{N,1}, \ldots , h_{N,M}, \kappa_0, \ldots ,\kappa_{i-1},\right.}  \nonumber \\
  &   \left.   \kappa_{i+1}, \ldots , \kappa_{2 N -1}\right]^{\rm T} \!=\!\left[ \bd{\theta}_{\rm Ch}^{\rm T}, \bd{\theta}_{{\rm MI}, i}^{\rm T} \right]^{\rm T} \!\!\!\in {\cal C}^{N M +2 N -1},\label{theta} \end{eqnarray}
whereby $N' \df N M +2 N -1$, and where  $h_{l,s} \in {\ho C}$ are complex Gaussian r.v.s of unit variance and zero mean, modeling independent flat fading coefficients from transmit antenna $l$ to receive antenna $s$.

Thereby, when transmitting (and detecting!) a $q_\cup$-ary r.v.\  $\kappa$ on a random position $i$ in an $N$-antenna label (see Fig.~\ref{parchann}), the channel state can be viewed as a random parameter with (random) realizations $\bd{\theta}_i$ that depend on the realizations of \sh.

\subsection{Coherent vs.\ non-coherent detection---$p_{\bd{\theta}}(\bd{y}|\bd{x})$ vs.\
$p(\bd{y}|\bd{x})$ }
 \label{cohvsnon}

The important implication of the above form of the state vector is
that  even when the receiver knows the MIMO channel perfectly,
lack of information about any interfering coordinates renders the
problem noncoherent (partial channel information, or no channel
state information)\footnote{unless some form of interference
cancellation is attempted}. The following discussion serves to
meaningfully characterize the differences between coherent,
noncoherent, and partially coherent scenarios.

The assumption of coherence vs.\ noncoherence essentially translates in whether the relevant pdf is conditioned on the channel state, or averaged over the channel state. One  characteristic of the MIMO channel is that the part of the channel state vector that depends of the multiple transmit antenna interference is always unknown (except possibly when talking about pilot symbols), and therefore the detection is always at least partially non-coherent; it will be necessary to average the pdf over the unknown part of the channel (even when the actual fading coefficients are known to the receiver), and this task will be naturally achieved by computing marginal probabilities. Note that the non-coherent aspect is responsible for the gap between CM and BICM in MIMO channels. Formally, this can be briefly described as follows.

In the coherent case the channel is characterized by the family of transition probability density functions
\be
\left\{p_{\bd{\theta}}(\bd{y}|\bd{x}) \left| \bd{\theta}\in {\ho C}^{N'}, \bd{x} \in {\ho R}^{N_1}, \bd{y}\in {\ho C}^{N_2} \right. \right\}
\ee
where the vector parameter $\bd{\theta}$ represents the channel state. The sequence of channel state vectors from a set of channel uses is denoted $\underline{\bd{\theta}}$. The essential assumption is  that $\bd{\theta}$ is independent of the channel input $\bd{x}$; e.g., in any of the parallel channels in Fig.~\ref{parchann} $\bd{\theta}_{{\rm MI},i} $ does not depend on $\kappa_i$---clearly $\bd{\theta}_{\rm Ch}$ is independent of $\kappa_i$, which makes $\bd{\theta}_i$ independent of $\kappa_i$. In addition it is assumed that, conditioned on $\underline{\bd{\theta}}$, the channel is memoryless, i.e.
\be
\label{memlesscoh}
p_{\underline{\bd{\theta}}}(\underline{\bd{y}}\left|\underline{\bd{x}}\right.)=\prod_k p_{\bd{\theta}_k}(\bd{y}_k\left|\bd{x}_k\right.).
\ee
Note that in ISI channels the state $\bd{\theta}$ depends on the input sequence, and therefore, given $\underline{\bd{\theta}}$, $\underline{\bd{y}}$ does depend on the input sequence (i.e.\ $\bd{y}_k$ depends not only on $\bd{x}_k$ but also on $\bd{x}_{k-1}$, etc.). Moreover, $\underline{\bd{\theta}}$ is assumed to be stationary and have finite memory. OFDM systems do verify the above assumptions, as the channel experienced by the encoder is flat (albeit correlated).

In the general noncoherent case lack of knowledge about $\underline{\bd{\theta}}$ prevents  the channel {\em transitions} from being memoryless. Nevertheless, under ideal interleaving, and for all finite index sets wherein interleaving renders the channel realizations independent, $E_{\underline{\bd{\theta}}}\{p_{\underline{\bd{\theta}}}(\underline{\bd{y}}\left|\underline{\bd{x}}\right.)\}=E_{\underline{\bd{\theta}}}\{\prod_{k\in {\cal K}} p_{\bd{\theta}_k}(\bd{y}_k\left|\bd{x}_k\right.)\}$ via \rf{memlesscoh}; because of the independence between the elements of the sequence $\underline{\bd{\theta}}$ the average of the product equals the product of averages, and one can define
\[
p(\bd{y}|\bd{x})\df E_{\underline{\bd{\theta}}}\{p_{\underline{\bd{\theta}}}(\underline{\bd{y}}\left|\underline{\bd{x}}\right.)\} = \prod_{k\in {\cal K}} p(\bd{y}_k\left|\bd{x}_k\right.)
\]
where a new average transition pdf is defined as
\be \label{avtran}
p(\bd{y}\left|\bd{x}\right.)=E_{\bd{\theta}}\{p_{\bd{\theta}}(\bd{y}\left|\bd{x}\right.)\}.
\ee
Using $p(\bd{y}\left|\bd{x}\right.)$ instead of $p_{\bd{\theta}}(\bd{y}\left|\bd{x}\right.)$ will allow transforming  metrics, decision rules, and expressions for average mutual information in the coherent case into same for noncoherent scenarios.

Finally, when relying on the model in Fig.\ \ref{parchann}, the
multiple input part of the state vector $\bd{\theta}_{{\rm MI},i}
$ will be unknown even when $\bd{\theta}_{{\rm Ch},i}$ is known;
averaging it out to obtain $p_{\bd{\theta}_{{\rm
Ch},i}}(\bd{y}\left|\bd{x}\right.)$ simply means computing
marginal probabilities for the coordinate of interest $\kappa_i$.
This will be used in Section~\ref{dermutinf}, and in producing the
numerical results.

\subsection{Derivation of mutual information
for CI}\label{dermutinf}

In order to compute the mutual information when transmitting a
$q_\cup$-ary r.v.\  on a random position in an $N$-antenna label,
consider a $q_\cup$-ary r.v.\ $\kappa$; the r.v.\  \sh, whose
outcome determines the switch position, is uniformly distributed
over $\{0, \ldots , 2N-1\}$, and has  known realizations $\sh =i$.
For the coherent case, where $\bd{\theta}$ and the permutation
pattern (i.e.\ \sh) are perfectly known to the receiver, one must
calculate the average mutual information $\overline{I(\kappa;
\bd{y}, \bd{\theta}, \sh)}$ as an average of $I(\kappa; \bd{y},
\bd{\theta}, \sh)$ over \sh, i.e.\ \be \overline{I(\kappa; \bd{y},
\bd{\theta}, \sh)} =\frac{1}{2 N}\sum_{i=0}^{2 N-1}I(\kappa;
\bd{y}, \bd{\theta}_i, \sh =i); \label{avcoh}\ee as noted, the
channel state can be viewed as a known random parameter with
(random) realizations $\bd{\theta}_i$; likewise, \sh\ is  a known
random parameter. Similarly, in the noncoherent case, one must
compute the average mutual information $\overline{I(\kappa;
\bd{y}, \sh)}$ as an average of $I(\kappa; \bd{y}, \sh)$ over \sh,
i.e.\ \be \overline{I(\kappa; \bd{y}, \sh)} =\frac{1}{2
N}\sum_{i=0}^{2 N-1}I(\kappa; \bd{y}, \sh =i); \label{avnoncoh}\ee
and, in the partially coherent case---when $\bd{\theta}_{\rm Ch}$
is known but the multiple input component of the state vector
$\bd{\theta}_{\rm MI}$ is unknown---the mutual information is
$\overline{I(\kappa; \bd{y}, \bd{\theta}_{\rm Ch}, \sh)}$ and must
be computed as an average of $I(\kappa; \bd{y}, \bd{\theta}_{\rm
Ch}, \sh)$ over \sh, i.e.\ \be \overline{I(\kappa; \bd{y},
\bd{\theta}_{\rm Ch}, \sh)} =\frac{1}{2 N}\sum_{i=0}^{2
N-1}I(\kappa; \bd{y}, \bd{\theta}_{\rm Ch}, \sh =i).
\label{avparcoh}\ee The third case is more physically meaningful
than the first (which assumes that the receiver always knows the
symbols transmitted from the competing transmit antennas); it is
the third mutual information that will be numerically computed in
the sequel.

Denoting by $\bd{\kappa} \df [\kappa_0, \ldots , \kappa_{2N-1}]^{\rm T}$ the complete MIMO channel input, the desired mutual information expressions in the above scenarios are
\begin{eqnarray} \overline{I(\bd{\kappa}; \bd{y}, \bd{\theta}, \sh)}= 2 N \overline{I(\kappa; \bd{y}, \bd{\theta}, \sh)}, \label{coh}\\
\overline{I(\bd{\kappa}; \bd{y}, \sh)}= 2 N \overline{I(\kappa; \bd{y}, \sh)}, \label{noncoh}\\
\overline{I(\bd{\kappa}; \bd{y}, \bd{\theta}_{\rm Ch}, \sh)}= 2 N \overline{I(\kappa; \bd{y}, \bd{\theta}_{\rm Ch}, \sh)}. \label{parcoh}\end{eqnarray}
In the sequel the subscript $i$ of $\bd{\theta}_i$ will be omitted for simplicity, but is to be understood in the theoretical, perfectly coherent case.

Consider first the coherent case, which requires computing $I(\kappa; \bd{y}, \bd{\theta}, \sh =i)$ via \rf{avcoh}.
By the chain rule for mutual information, $I(\kappa; \bd{y}, \bd{\theta}, \sh =i)=I(\kappa; \sh = i) + I(\kappa; \bd{y}, \bd{\theta}|\sh =i)\stackrel{\mbox{\scriptsize (a)}}{=}I(\kappa; \bd{\theta}|\sh =i)+I(\kappa; \bd{y}|\bd{\theta}, \sh =i)\stackrel{\mbox{\scriptsize (b)}}{=}I(\kappa; \bd{y}|\bd{\theta}, \sh =i)$; (a) follows because $I(\kappa; \sh =i)=0$ due to the independence between $\kappa$ and $\sh =i$, and  similarly (b). It suffices then to average over the conditional pdf $p(\kappa, \bd{y}, \bd{\theta}|\sh =i)$, and---since $ I(\kappa; \bd{y}, \bd{\theta}|\sh =i)=I(\kappa; \bd{y}|\bd{\theta}, \sh =i)$---the conditional mutual information of $\kappa$, $\bd{y}$,  given $\bd{\theta}$ and $\sh =i$, for the case when $\kappa$ is uniformly distributed in ${\cal Q}_\cup$, reduces   to (see Appendix~\ref{app1})

\be I(\kappa; \bd{y}|\bd{\theta}, \sh =i)=\log_2 (q_\cup) - E_{\kappa, \bd{y}, \bd{\theta}} \left[ \log_2 \frac{\sum_{\bd{z} \in {\cal X}}p_{\bd{\theta}}(\bd{y}|\bd{z})}{\sum_{\bd{\zeta} \in {\cal X}_\kappa^i}p_{\bd{\theta}}(\bd{y}|\bd{\zeta})}\right]\label{micohconds}\ee
where $\kappa$, $\bd{y}$,  and $\bd{\theta}$ are conditionally distributed as (see Appendix~\ref{app2})
\be
p(\kappa, \bd{y}, \bd{\theta}|\sh =i) = q_{\cup}^{-2 N}p(\bd{\theta})\sum_{\bd{\zeta} \in {\cal X}_\kappa^i} p_{\bd{\theta}}(\bd{y}|\bd{\zeta} ).
\label{pdfcohconds} \ee
Above, ${\cal X}_\kappa^i$ denotes the set of all $N$-antenna labels having $\kappa$ on position $i$, $i=0, \ldots , 2N-1$. Finally, using eqs.\ \rf{avcoh}, \rf{coh}, one obtains the mutual information with CI in the coherent scenario, for the case when $\kappa$ is uniformly distributed in ${\cal Q}_\cup$,  to be
\be \overline{I(\bd{\kappa}; \bd{y}, \bd{\theta}, \sh)}=2 N \log_2 (q_\cup) - \sum_{i=0}^{2 N-1} E_{\kappa, \bd{y}, \bd{\theta}} \left[ \log_2 \frac{\sum_{\bd{z} \in {\cal X}}p_{\bd{\theta}}(\bd{y}|\bd{z})}{\sum_{\bd{\zeta} \in {\cal X}_\kappa^i}p_{\bd{\theta}}(\bd{y}|\bd{\zeta})}\right].\label{micoh}\ee

Likewise, in  the case when $\kappa$ is non-uniformly distributed in ${\cal Q}_\cup$,
\be I(\kappa; \bd{y}|\bd{\theta}, \sh =i)= - E_{\kappa, \bd{y}, \bd{\theta}} \left[ \log_2 \frac{\sum_{\bd{z} \in {\cal X}}p_{\bd{\theta}}(\bd{y}|\bd{z}) p(\bd{z})}{\sum_{\bd{\zeta} \in {\cal X}_\kappa^i}p_{\bd{\theta}}(\bd{y}|\bd{\zeta})p(\bd{\zeta})}\right],\label{micohcondsnonunif}\ee

\be
p(\kappa, \bd{y}, \bd{\theta}|\sh =i) = p(\kappa)p(\bd{\theta})\sum_{\bd{\zeta} \in {\cal X}_\kappa^i} p_{\bd{\theta}}(\bd{y}|\bd{\zeta} )p(\bd{\zeta}).
\label{pdfcohcondsnonunif} \ee

\be \overline{I(\bd{\kappa}; \bd{y}, \bd{\theta}, \sh)}= - \sum_{i=0}^{2 N-1} E_{\kappa, \bd{y}, \bd{\theta}} \left[ \log_2 \frac{\sum_{\bd{z} \in {\cal X}}p_{\bd{\theta}}(\bd{y}|\bd{z})p(\bd{z})}{\sum_{\bd{\zeta} \in {\cal X}_\kappa^i}p_{\bd{\theta}}(\bd{y}|\bd{\zeta})p(\bd{\zeta})}\right].\label{micohnonunif}\ee

Similarly in the noncoherent case, when $\kappa$ is uniformly distributed in ${\cal Q}_\cup$,
\be \overline{I(\bd{\kappa}, \bd{y}, \sh)}=2 N \log_2 (q_\cup) - \sum_{i=0}^{2 N-1} E_{\kappa, \bd{y}} \left[ \log_2 \frac{\sum_{\bd{z} \in {\cal X}}p(\bd{y}|\bd{z})}{\sum_{\bd{\zeta} \in {\cal X}_\kappa^i}p(\bd{y}|\bd{\zeta})}\right],\label{minoncoh}\ee
where $p(\bd{y}|\bd{z})$ is obtained as in Section \ref{cohvsnon}; and in the partial coherent case, when the fading coefficients are known to the receiver,
\be \overline{I(\bd{\kappa}; \bd{y}, \bd{\theta}_{\rm Ch}, \sh)}=2 N \log_2 (q_\cup) - \sum_{i=0}^{2 N-1} E_{\kappa, \bd{y}, \bd{\theta}_{\rm Ch}} \left[ \log_2 \frac{\sum_{\bd{z} \in {\cal X}}p_{\bd{\theta}_{\rm Ch}}(\bd{y}|\bd{z})}{\sum_{\bd{\zeta} \in {\cal X}_\kappa^i}p_{\bd{\theta}_{\rm Ch}}(\bd{y}|\bd{\zeta})}\right],\label{parcoh}\ee
where averaging over the multiple input component of the channel state vector, per \rf{avtran}, is implemented naturally via marginal probabilities.

Similar expressions can be derived in the noncoherent and partially coherent scenarios for the case when $\kappa$ is non-uniformly distributed in ${\cal Q}_\cup$; in the latter case
\be \overline{I(\bd{\kappa}; \bd{y}, \bd{\theta}_{\rm Ch}, \sh)}= - \sum_{i=0}^{2 N-1} E_{\kappa, \bd{y}, \bd{\theta}_{\rm Ch}} \left[ \log_2 \frac{\sum_{\bd{z} \in {\cal X}}p_{\bd{\theta}_{\rm Ch}}(\bd{y}|\bd{z})p(\bd{z})}{\sum_{\bd{\zeta} \in {\cal X}_\kappa^i}p_{\bd{\theta}_{\rm Ch}}(\bd{y}|\bd{\zeta})p(\bd{\zeta})}\right],\label{parcohnonunif}\ee

\section{Numerical results}
\label{mutinfcoordinterlnumres} For a variety of MIMO
configurations and constellation sizes the mutual information in
the presence of CI is shown in Figs.\
\ref{fig:16qam2x1}--\ref{fig:nqamnx2}, along with mutual
information for BICM and CM. While the fading coefficients are
assumed known to the receiver, the interfering symbols transmitted
from other antennas are not known, and are averaged out as
discussed in Section \ref{cohvsnon} (essentially computing
marginal likelihood probabilities); mathematically, the figures
plot $\overline{I}(\bd{\kappa}; \bd{y}, \bd{\theta}_{\rm Ch})$.

\subsection{Complex constellations that are invariant to CI} \label{constellinvCI}

\begin{figure}[htbp] \begin{center} \epsfig{file=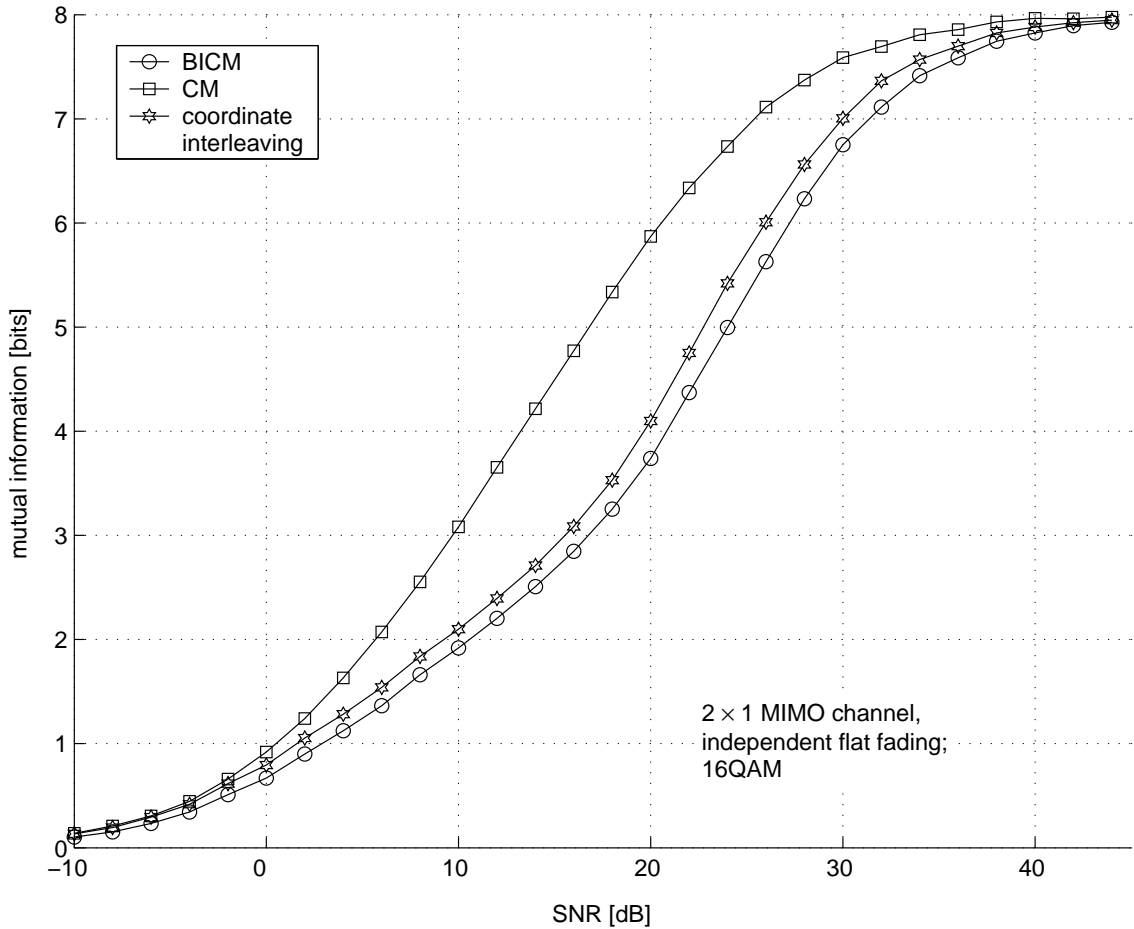, width=1\textwidth} \caption{Mutual information comparison between BICM (with pure Gray mapping) and CM with and without CI, in a Rayleigh fading MIMO configuration with two  transmit and one receive antennas, and 16QAM constellations.}\label{fig:16qam2x1} \end{center} \end{figure}

\begin{figure}[htbp] \begin{center} \epsfig{file=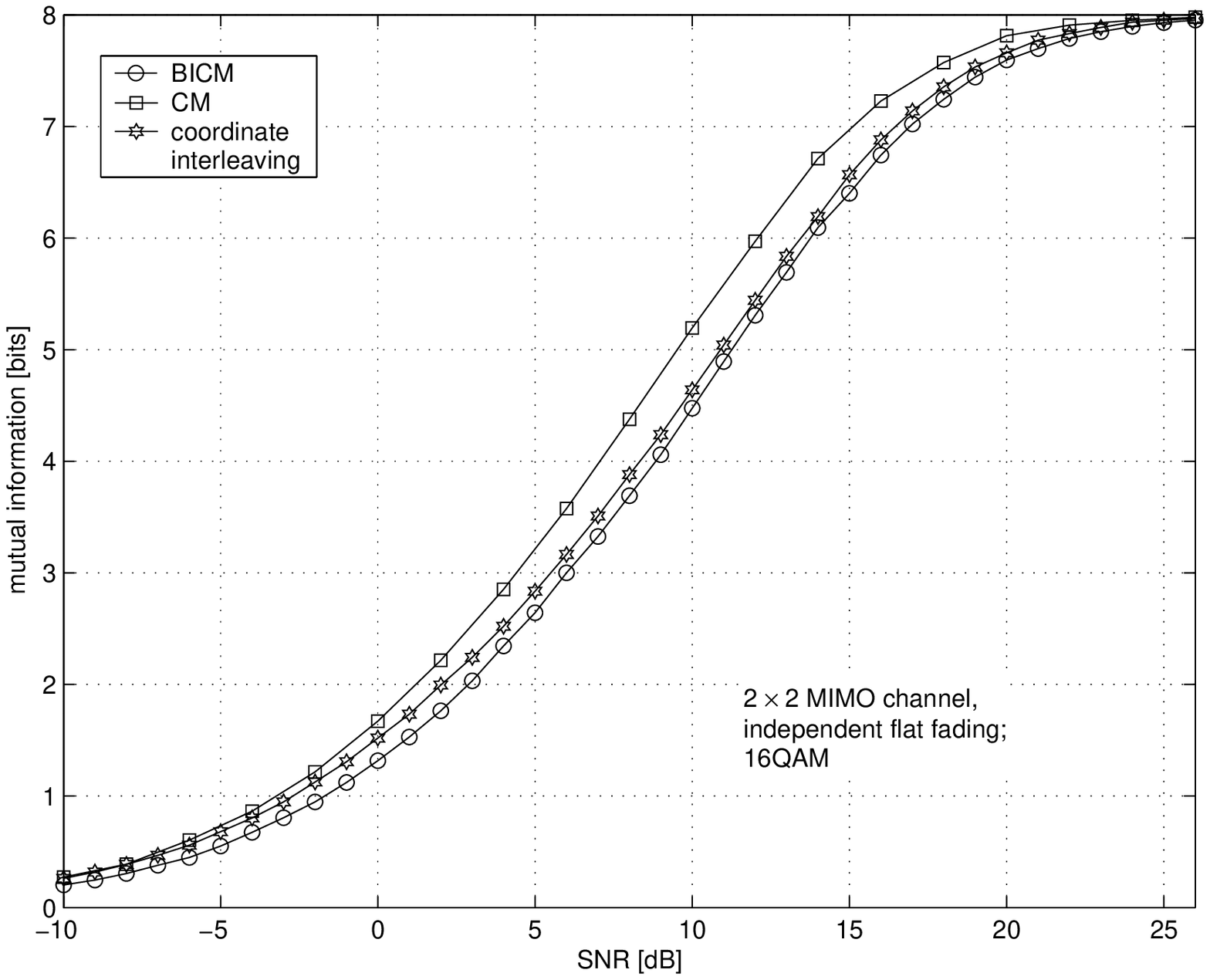, width=1\textwidth} \caption{Mutual information comparison between BICM (with pure Gray mapping) and CM with and without CI, in a Rayleigh fading MIMO configuration with two  transmit and  receive antennas, and 16QAM constellations.}\label{fig:16qam2x2} \end{center} \end{figure}

\begin{figure}[htbp] \begin{center} \epsfig{file=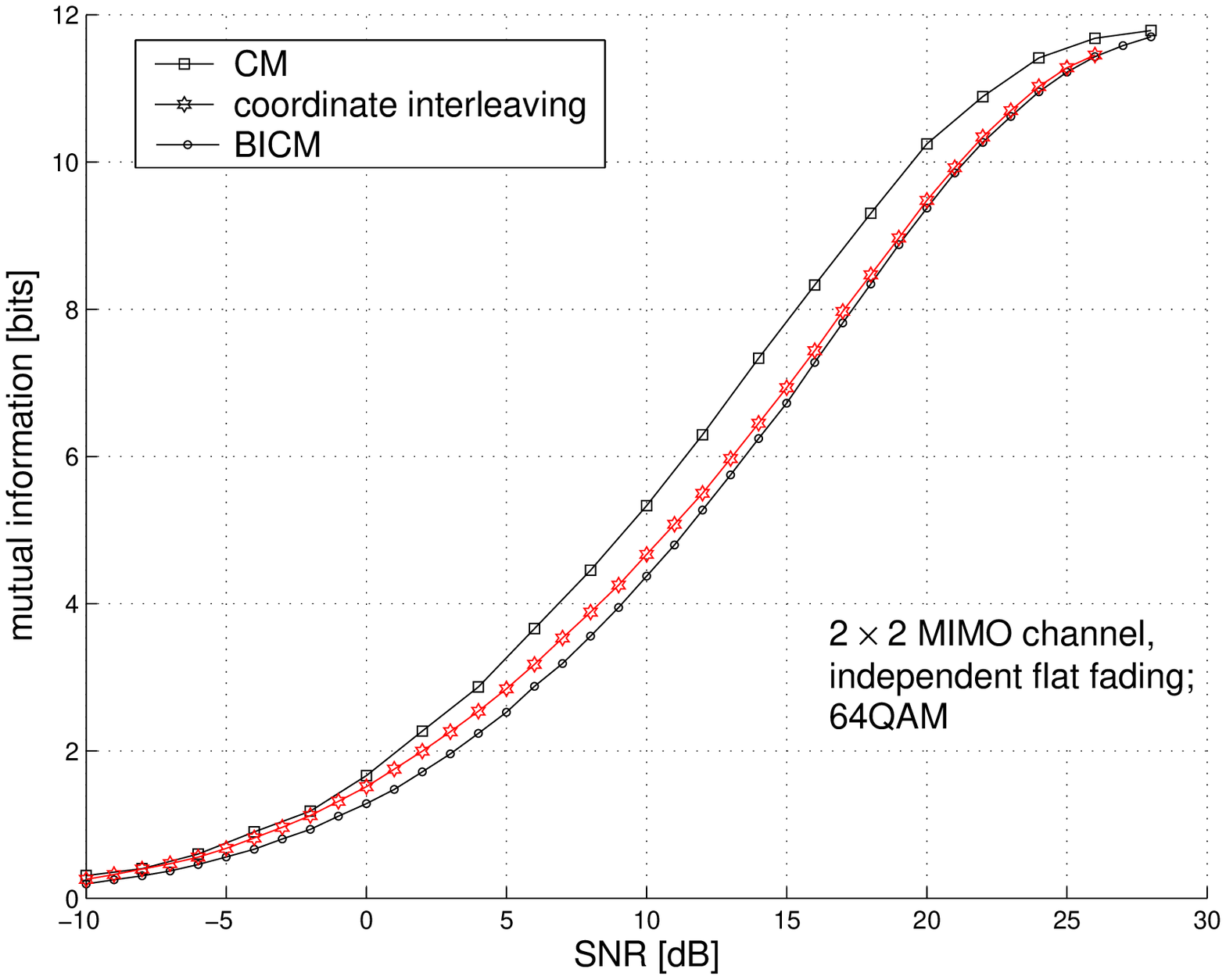, width=1\textwidth} \caption{Mutual information comparison between BICM (with pure Gray mapping) and CM with and without CI, in a Rayleigh fading MIMO configuration with two  transmit and  receive antennas, and 64QAM constellations.}\label{fig:64qam2x2} \end{center} \end{figure}

\begin{figure}[htbp] \begin{center} \epsfig{file=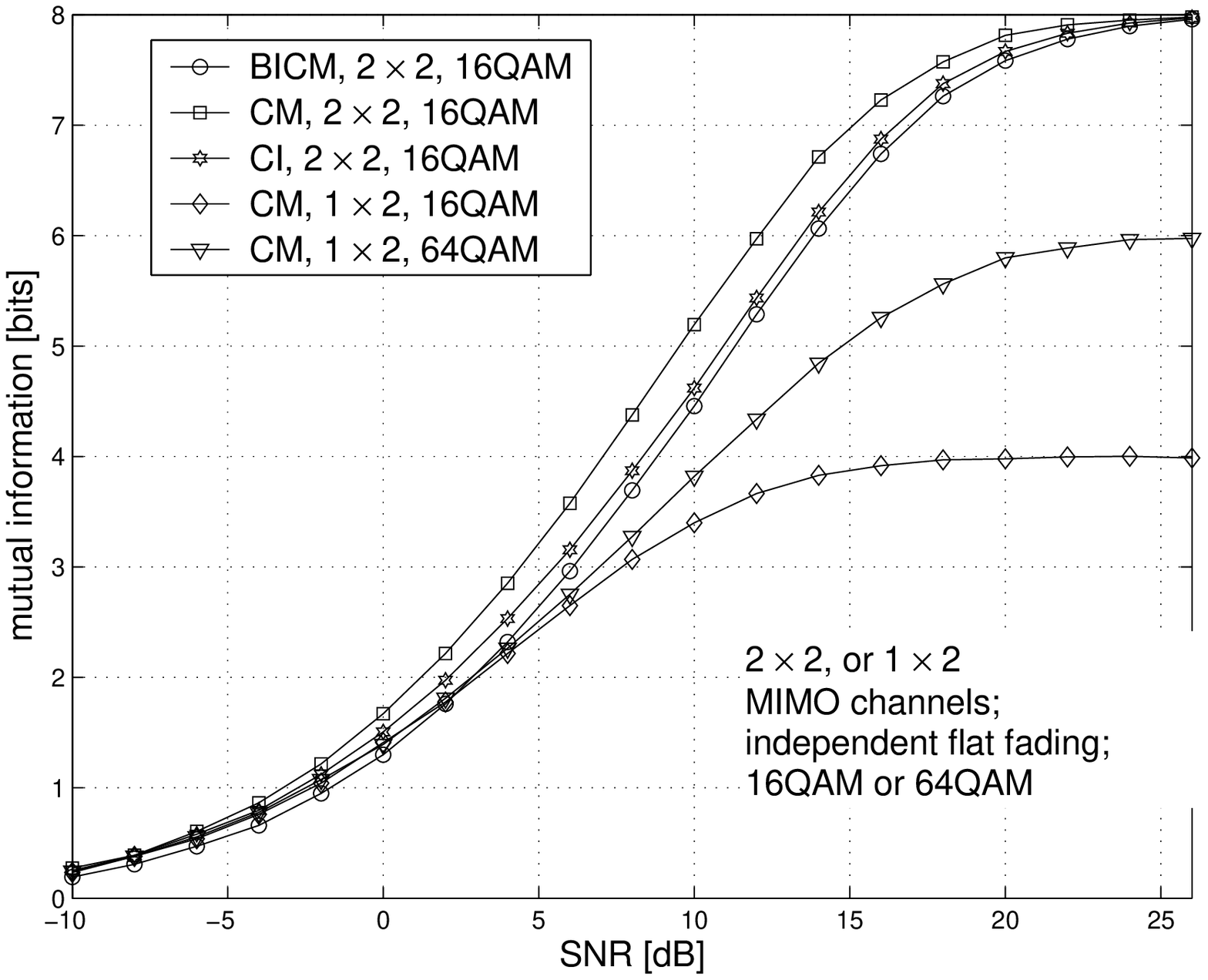, width=1\textwidth} \caption{Mutual information comparison between BICM (with pure Gray mapping) and CM with and without CI, in a Rayleigh fading MIMO configuration with either one or two  transmit,  and  receive antennas, and 16QAM, 64QAM constellations. Interestingly, {\em while the $2 \times 2$ BICM curve crosses below the $1 \times 2$ curves at low SNR, the $2 \times 2$ CI  curve is always above the $1 \times 2$ curves}.}\label{fig:nqamnx2} \end{center} \end{figure}

Complex constellations like 4PSK, 16QAM, 64QAM are invariant to CI; i.e., all possible combinations of valid coordinates of either type (real or imaginary) result in valid constellation points. Mathematically this means that ${\cal Q}= ({\cal Q}_I \cup {\cal Q}_Q) \times ({\cal Q}_I \cup {\cal Q}_Q)$, as opposed to ${\cal Q}\subset ({\cal Q}_I \cup {\cal Q}_Q) \times ({\cal Q}_I \cup {\cal Q}_Q)$. One can easily visualize this, and such  cases are illustrated in Figs.\ \ref{fig:16qam2x1}, \ref{fig:16qam2x2}, \ref{fig:64qam2x2}; Fig.\ \ref{fig:nqamnx2} includes two  single transmit antenna scenarios for comparison.

\subsection{Complex constellations that grow in size after CI}
\label{constellnoninvCI}

An example of complex constellation that is {\em not} invariant
with respect to CI is the cross-shaped 32QAM. This constellation
does not admit a pure Gray mapping, but there exists an optimum
impure Gray coding \cite{smi:odd} with a Gray penalty of 7/6 (this
should be as close to 1 as possible; pure Gray mapping corresponds
to Gray penalty of 1). More importantly, ${\cal Q}\subset ({\cal
Q}_I \cup {\cal Q}_Q) \times ({\cal Q}_I \cup {\cal Q}_Q)$, and
thereby putting together arbitrary  coordinates---which are
otherwise valid---results in points that are not necessarily
present in the original constellation; the richer constellation
can send more bits per channel use asymptotically with SNR, which
means that the mutual information curve in the presence of CI can
cross even the CM curve at high mutual information values; see
Fig.~\rf{fig:32qam2x2}. The implication is that {\em it is
possible to alleviate the capacity limit by CI, at high
information levels}.

\begin{figure}[htbp] \begin{center} \epsfig{file=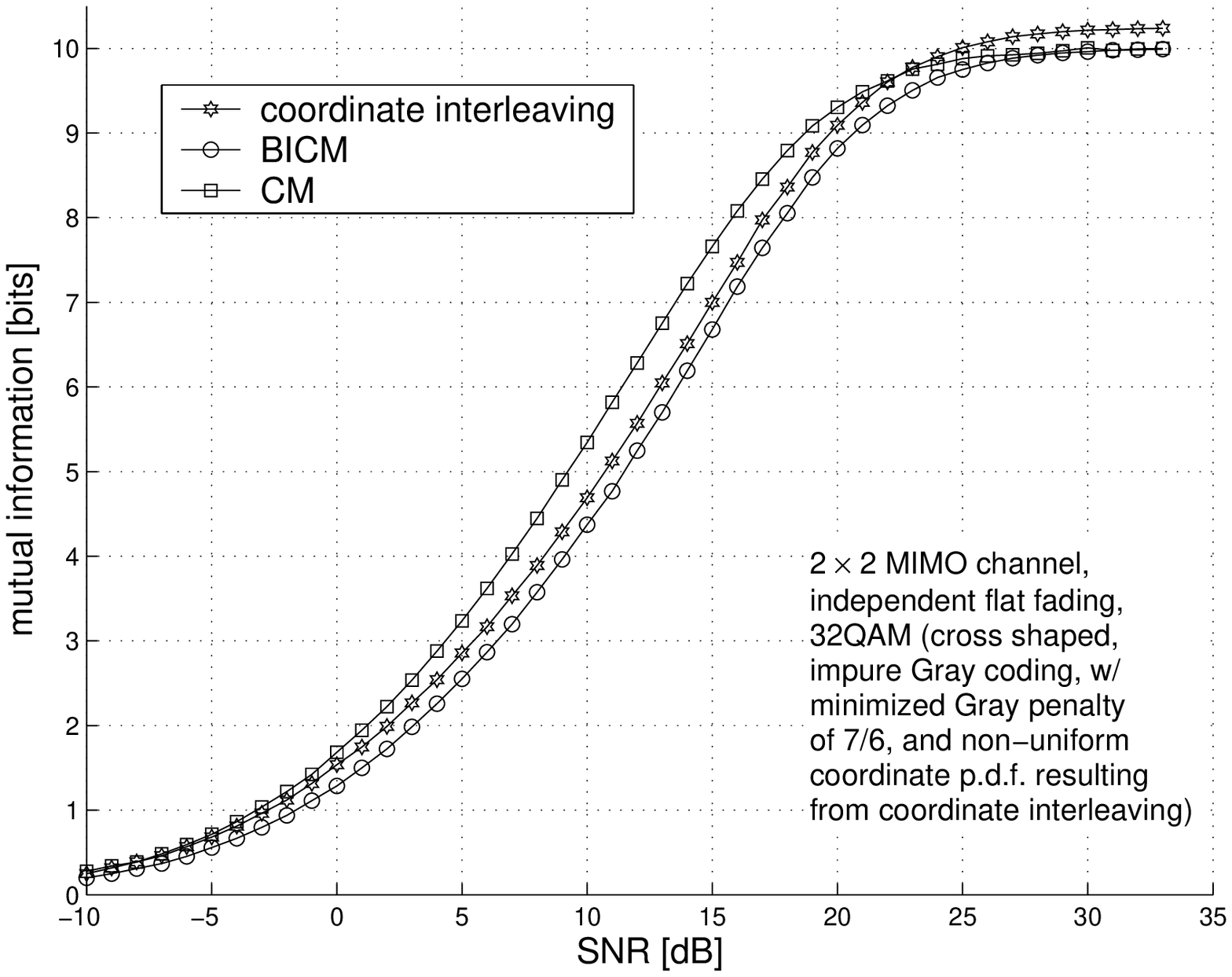, width=1\textwidth} \caption{Mutual information comparison between BICM and CM with and without CI, in a Rayleigh fading MIMO configuration with two  transmit and  receive antennas, and 32QAM constellations.}\label{fig:32qam2x2} \end{center} \end{figure}

\begin{figure}[htbp] \begin{center} \epsfig{file=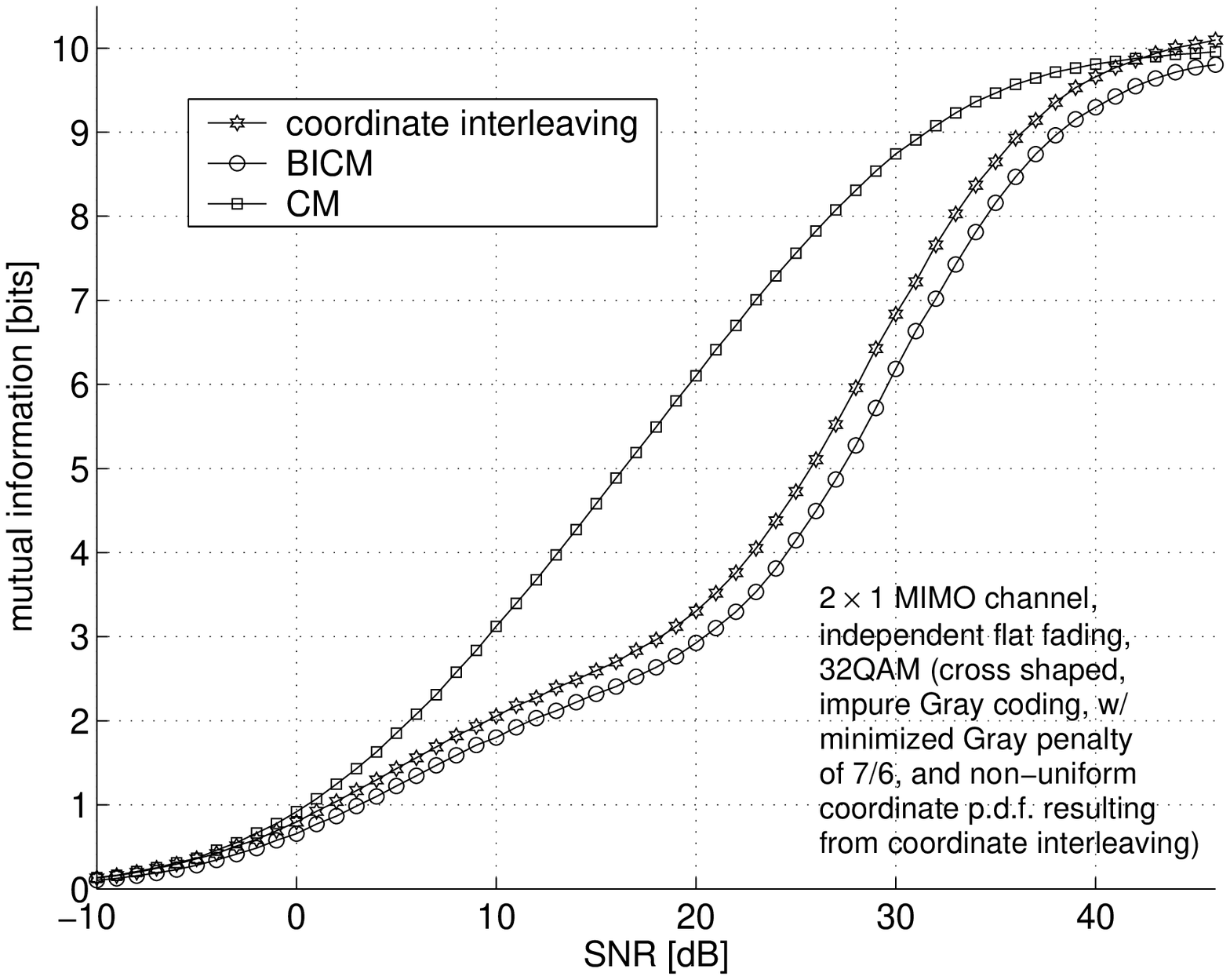, width=1\textwidth} \caption{Mutual information comparison between BICM and CM with and without CI, in a Rayleigh fading MIMO configuration with two  transmit and one receive antennas, and 32QAM constellations.}\label{fig:32qam2x1} \end{center} \end{figure}

It is possible to evaluate exactly the upper mutual information margin for a given, finite, constellation size; the simple observation is that for any two r.v.s $X, Y$
\[I(X;Y)=H(X)- H(X|Y);\]
note that, asymptotically with SNR---i.e., as the noise becomes negligible---$Y$ resembles $X$ more and more, and replicates  $X$ in the limit. Thereby, as ${\rm SNR} \rightarrow \infty$, $H(X|Y)\rightarrow H(X|X)= 0$  and  $I(X;Y)\rightarrow H(X)$; this was to be expected, since the maximum mutual information cannot exceed the source's entropy. In this context $X$ will be a r.v.\ with realizations in $\cal M$ from \rf{perantciconstell}. A cross shaped 32QAM constellation $\cal Q$, when  used on each transmit antenna prior to CI, gives rise to  a thirty six point constellation $\cal M$ after CI (replicated $N$ times to form $\cal X$ when $N$ transmit antennas are used). In the enhanced constellation $\cal M$ there are four points that occur with probability of $1/64$, sixteen points with probability of $3/128$, and sixteen points with probability of $9/256$. The set of probabilities of complex $N$-tuples corresponding to  the $N$ transmit antennas is the $N$ fold Cartesian product of this set of probabilities with itself; alternatively, the entropy of the $N$ antenna labels is $N$ times the entropy corresponding to one antenna (antenna streams are assumed independent). It can be easily verified that for the thirty six point enhanced constellation the entropy equals $5.12$ bits, which after multiplication by $N=2$ yields more than the $10$ bit information limit that bounds both CM and BICM using the original cross shaped 32QAM constellation; this is indeed consistent with Figs.~\ref{fig:32qam2x2}, \ref{fig:32qam2x1}, and explains the cross over between CM and CI.

\begin{figure}[htbp] \begin{center} \epsfig{file=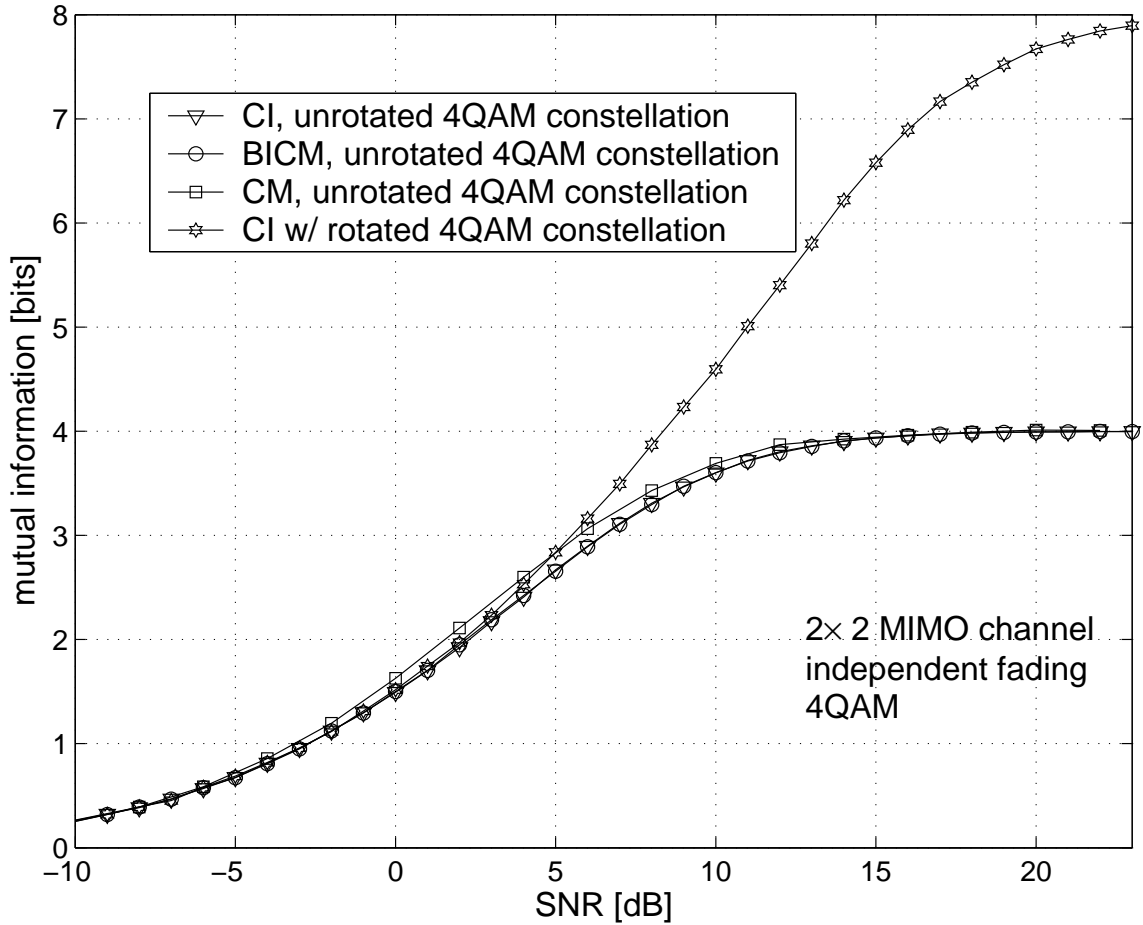, width=1\textwidth} \caption{Mutual information comparison between BICM and CM with and without CI, in a Rayleigh fading MIMO configuration with two  transmit and receive antennas, using a 4QAM  constellation with and without rotation.}\label{fig:4qamWrotat2x2} \end{center} \end{figure}

A more pronounced constellation enhancement effect  can be
obtained by constellation rotation; the extent of constellation
enrichment after CI will be more dramatic than in the cross-shaped
32QAM case. Rotated constellations may possess  an advantage when
CI is used, and may  thereby be desirable, because CI results in
an enriched effective constellation; grouping scrambled
coordinates results in complex values that do not necessarily
belong to the original constellation, and the richer, effective
channel alphabet leads to higher mutual information
(asymptotically with SNR, see Fig.~\ref{fig:32qam2x2}), and {\em
even crosses above the CM curve in the range of interest}.
Fig.~\ref{fig:32qam2x2} is a very limited illustration because
constellation enrichment was achieved simply due to the inherent
structure of the cross-shaped 32QAM constellation (increase was
marginal, from 32 to 36 points). Deliberate rotations will
increase the constellation size to larger extents, and probably
cause the CI curve to cross the CM curve earlier.
Fig.~\ref{fig:4qamWrotat2x2} shows the effect of rotating a 4PSK
constellation by an angle that makes the enhanced constellation be
a scaled version of a 16QAM constellation. {\em It might be
possible to engage (activate) CI selectively, at hight SNR, in
order to increase the throughput while delaying increasing the
constellation size.}

This aspect deserves further consideration.

\begin{figure}[bt]
\scalebox{1.0}{\input{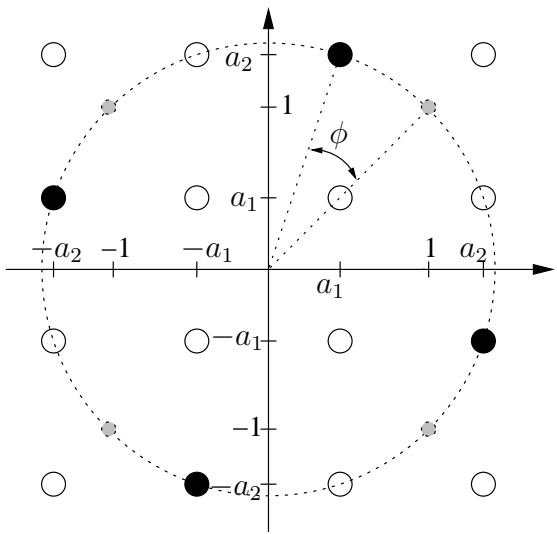}} \caption{When rotated
by $\phi = \pi/4 - \arctan (1/3) $ and coordinate interleaved, a
4PSK constellation $\{\pm 1 \pm j\}$ gives rise  to a 16QAM
constellation scaled by $\sqrt{2} \sin \left(\arctan(1/3)\right)
$.}\label{constellrot}
\end{figure}

\section{Conclusions} \label{concl}

This paper identifies a shortcoming of bit interleaved coded
modulation in MIMO fading channels, in terms of constrained
i.i.d.\ mutual information relative to coded modulation; while the
mutual information curves for coded modulation and bit interleaved
coded modulation overlap in the case of a single input single
output channel, a loss in mutual information was observed for bit
interleaved coded modulation in MIMO channels. An alternative to
bit interleaved coded modulation in MIMO channels, namely
coordinate interleaving, was proposed and analyzed. Coordinate
interleaving offers better diversity, and improves constrained
mutual information over the bit interleaved coded modulation.
Effects like constellation rotation and non-uniform constellation
points in the presence of CI are also studied.

\appendices
\section{}\label{app1}
The derivation of \rf{micohconds} is as follows. First, note that
referring to a $q_\cup$-ary r.v.\  while conditioning on $\sh = i$
is equivalent to referring to $\kappa_i$. By definition, if the
realizations of a r.v.\  $A$ are from $\cal A$, with cardinality
$|{\cal A}|$, then the mutual information between r.v.s $A,B$ is
\begin{eqnarray}
\lefteqn{ I(A; B) = E_{A, B}\left[ \log_2 \frac{p(a,b)}{p(a)p(b)} \right]} \nonumber \\
& &  =-E_{A, B} \left[ \log_2 \frac{p(b)}{p(b|a)}\right]\nonumber\\
& & =-E_{p(a,b)} \left[ \log_2 \frac{\sum_{\alpha \in {\cal A}}p(b|\alpha)p(\alpha)}{p(b|a)}\right] \label{a0}
\end{eqnarray}
If $A$ is uniformly distributed in $\cal A$ then $p(\alpha) = |{\cal A}|^{-1}, \forall \alpha \in {\cal A}$, which yields
\begin{eqnarray}
I(A; B) =    -\log_2 |{\cal A}|^{-1}- E_{p(a,b)} \left[ \log_2 \frac{\sum_{\alpha \in {\cal A}}p(b|\alpha)}{p(b|a)}\right], \label{a00}
\end{eqnarray} Similarly, if the realizations of a r.v.\  $A$ are
uniformly distributed in  $\cal A$, with $|{\cal A}|=2^m$, then
the conditional mutual information between r.v.s $A,B$,
conditioned on $C$ is $ I(A; B|C) = $ $  -\log_2 |{\cal A}|^{-1}-
E_{A, B, C} \left[ \log_2 \frac{\sum_{\alpha \in {\cal
A}}p(b|\alpha,c)}{p(b|a,c)}\right]\label{a1} $ $=  -\log_2 |{\cal
A}|^{-1}- E_{A, B, C} \left[ \log_2 \frac{\sum_{a \in {\cal
A}}p_c(b|a)}{p_c(b|a)}\right], \label{a1} $ where conditioning on
$c$ has been represented by writing $c$ as a subscript. Then, in
the case when $\kappa$ is uniformly distributed in ${\cal
Q}_\cup$, \be I(\kappa; \bd{y}|\bd{\theta}, \sh =i)= \log_2
(q_\cup)  - E_{\kappa, \bd{y}, \bd{\theta}} \left[ \log_2
\frac{\sum_{a \in {\cal Q}_\cup}p_{\bd{\theta}}(\bd{y}|\kappa = a,
\sh = i)}{p_{\bd{\theta}}(\bd{y}|\kappa, \sh =
i)}\right].\label{a2}\ee But  by denoting
$\bd{\kappa}_{\overline{i}}\df \{\kappa_0, \ldots, \kappa_{i-1},
\kappa_{i+1}, \ldots, \kappa_{2N -1} \}$ one can write
\begin{eqnarray}
\lefteqn{ p_{\bd{\theta}}(\bd{y}|\kappa, \sh = i) =  p_{\bd{\theta}}(\bd{y}|\kappa_i= \kappa ) } \nonumber \\
  & & = \sum_{\bd{\kappa}_{\overline{i}}}p_{\bd{\theta}}\left(\bd{y}, \bd{\kappa}_{\overline{i}}\ \ | \kappa_i= \kappa\right)\nonumber\\
 & & =  \sum_{\bd{\kappa}_{\overline{i}}}p_{\bd{\theta}}\left(\bd{y} |\ \ \kappa_i= \kappa,  \bd{\kappa}_{\overline{i}} \right) q_\cup^{-(2 N -1)}\nonumber \\
 & & =  q_\cup^{-(2 N -1)} \sum_{\bd{\zeta} \in {\cal X}_\kappa^i} p_{\bd{\theta}}(\bd{y}|\bd{\zeta} ), \nonumber
\end{eqnarray}
and
\begin{eqnarray}
\lefteqn{  \sum_{a \in {\cal Q}_\cup}p_{\bd{\theta}}(\bd{y}|\kappa = a, \sh = i)  }\nonumber\\
& & = \sum_{a \in {\cal Q}_\cup}p_{\bd{\theta}}(\bd{y}|\kappa_i = a)  \nonumber \\
 & & = \sum_{a \in {\cal Q}_\cup} \sum_{\bd{\kappa}_{\overline{i}}}p_{\bd{\theta}}\left(\bd{y}, \bd{\kappa}_{\overline{i}}\ \ | \kappa_i= a\right)\nonumber\\
 & & =  \sum_{a \in {\cal Q}_\cup}\sum_{\bd{\kappa}_{\overline{i}}}p_{\bd{\theta}}\left(\bd{y} |\ \ \kappa_i= a,  \bd{\kappa}_{\overline{i}} \right) q_\cup^{-(2 N -1)}\nonumber \\
& & =  q_\cup^{-(2 N -1)} \sum_{\bd{z} \in {\cal X}} p_{\bd{\theta}}(\bd{y}|\bd{z} ), \nonumber
\end{eqnarray}
hence \rf{micohconds}  follows after simplifying $q_{\cup}^{-(2 N -1)}$.
Likewise,  in the case when $\kappa$ is non-uniformly distributed in ${\cal Q}_\cup$,
\be I(\kappa; \bd{y}|\bd{\theta}, \sh =i)= - E_{\kappa, \bd{y}, \bd{\theta}} \left[ \log_2 \frac{\sum_{a \in {\cal Q}_\cup}p_{\bd{\theta}}(\bd{y}|\kappa = a, \sh = i)p(a)}{p_{\bd{\theta}}(\bd{y}|\kappa, \sh = i)}\right]\label{a2}\ee
By denoting $\bd{\kappa}_{\overline{i}}\df \{\kappa_0, \ldots, \kappa_{i-1}, \kappa_{i+1}, \ldots, \kappa_{2N -1} \}$ one can write
\begin{eqnarray}
\lefteqn{ p_{\bd{\theta}}(\bd{y}|\kappa, \sh = i) p(\kappa)=  p_{\bd{\theta}}(\bd{y}|\kappa_i= \kappa ) } \nonumber \\
  & & = \sum_{\bd{\kappa}_{\overline{i}}}p_{\bd{\theta}}\left(\bd{y}, \bd{\kappa}_{\overline{i}}\ \ | \kappa_i= \kappa\right)\nonumber\\
 & & =  \sum_{\bd{\kappa}_{\overline{i}}}p_{\bd{\theta}}\left(\bd{y} |\ \ \kappa_i= \kappa,  \bd{\kappa}_{\overline{i}} \right) p(\bd{\kappa}_{\overline{i}}) \nonumber \\
 & & =   \sum_{\bd{\zeta} \in {\cal X}_\kappa^i} p_{\bd{\theta}}(\bd{y}|\bd{\zeta} )p(\bd{\zeta}), \nonumber
\end{eqnarray}
and
\begin{eqnarray}
\lefteqn{  \sum_{a \in {\cal Q}_\cup}p_{\bd{\theta}}(\bd{y}|\kappa = a, \sh = i) p(a) }\nonumber\\
& & = \sum_{a \in {\cal Q}_\cup}p_{\bd{\theta}}(\bd{y}|\kappa_i = a)p(a) \nonumber \\
 & & = \sum_{a \in {\cal Q}_\cup} \sum_{\bd{\kappa}_{\overline{i}}}p_{\bd{\theta}}\left(\bd{y}, \bd{\kappa}_{\overline{i}}\ \ | \kappa_i= a\right)p(a)\nonumber\\
 & & =  \sum_{a \in {\cal Q}_\cup}\sum_{\bd{\kappa}_{\overline{i}}}p_{\bd{\theta}}\left(\bd{y} |\ \ \kappa_i= a,  \bd{\kappa}_{\overline{i}} \right) p(\bd{\kappa}_{\overline{i}})p(a) \nonumber \\
& & =  \sum_{\bd{z} \in {\cal X}} p_{\bd{\theta}}(\bd{y}|\bd{z} ) p({\bd z}), \nonumber
\end{eqnarray}
hence \rf{micohcondsnonunif}  follows.
\section{}\label{app2}
The derivation of \rf{pdfcohconds} is as follows. Note that  $\kappa, \bd{\theta}$ do not depend on \sh; then
\begin{eqnarray}
\lefteqn{p(\kappa, \bd{y}, \bd{\theta}|\sh =i) =p(\bd{y}|\kappa, \bd{\theta}, \sh =i) p(\kappa, \bd{\theta}|\sh =i)}\nonumber\\
& & =p_{\bd{\theta}}(\bd{y}|\kappa_i = \kappa) p(\kappa, \bd{\theta}|\sh =i)\nonumber\\
& & =p_{\bd{\theta}}(\bd{y}|\kappa_i = \kappa) p(\kappa) p(\bd{\theta})\nonumber\\
& & =q_\cup^{-1}p(\bd{\theta}) \sum_{\bd{\kappa}_{\overline{i}}}p_{\bd{\theta}}(\bd{y}, \bd{\kappa}_{\overline{i}}|\kappa_i = \kappa)\nonumber \\
& & =q_\cup^{-1}p(\bd{\theta}) \sum_{\bd{\kappa}_{\overline{i}}}p_{\bd{\theta}}(\bd{y}|\bd{\kappa}_{\overline{i}}, \kappa_i = \kappa)q_{\cup}^{-(2 N -1)}\nonumber \\
& & =q_{\cup}^{-2 N}p(\bd{\theta})\sum_{\bd{\zeta} \in {\cal
X}_\kappa^i} p_{\bd{\theta}}(\bd{y}|\bd{\zeta} ). \nonumber
\nonumber\end{eqnarray}

Finally, when $\kappa$ is non-uniformly distributed in ${\cal Q}_\cup$, \rf{pdfcohcondsnonunif} is proved via
\begin{eqnarray}
\lefteqn{p(\kappa, \bd{y}, \bd{\theta}|\sh =i) =p(\bd{y}|\kappa, \bd{\theta}, \sh =i) p(\kappa, \bd{\theta}|\sh =i)}\nonumber\\
& & =p_{\bd{\theta}}(\bd{y}|\kappa_i = \kappa) p(\kappa, \bd{\theta}|\sh =i)\nonumber\\
& & =p_{\bd{\theta}}(\bd{y}|\kappa_i = \kappa) p(\kappa) p(\bd{\theta})\nonumber\\
& & =p(\bd{\theta}) \sum_{\bd{\kappa}_{\overline{i}}}p_{\bd{\theta}}(\bd{y}, \bd{\kappa}_{\overline{i}}|\kappa_i = \kappa) p(\kappa)\nonumber \\
& & =p(\kappa) p(\bd{\theta}) \sum_{\bd{\kappa}_{\overline{i}}}p_{\bd{\theta}}(\bd{y}|\bd{\kappa}_{\overline{i}}, \kappa_i = \kappa)p(\bd{\kappa}_{\overline{i}}) \nonumber \\
& & =p(\kappa) p(\bd{\theta})\sum_{\bd{\zeta} \in {\cal
X}_\kappa^i} p_{\bd{\theta}}(\bd{y}|\bd{\zeta} ) p(\bd{\zeta}).
\nonumber \nonumber\end{eqnarray}

\pagebreak

\end{document}